\documentclass{jpp}
\usepackage{graphicx}

\usepackage[utf8]{inputenc}
\usepackage[T1]{fontenc}
\usepackage{amsmath}

\shorttitle{Plasma rotation driven by lasers with zero angular momentum}
\shortauthor{C. Willim, T. Silva, L. O. Silva and J. Vieira}

\title{Plasma rotation driven by lasers with zero angular momentum}

\author{Camilla Willim \aff{1}
  \corresp{\email{camilla.willim@tecnico.ulisboa.pt}},
  Thales Silva \aff{1}
  Lu\'is O. Silva \aff{1}
 \and Jorge Vieira \aff{1}}

\affiliation{%
 GoLP/Instituto de Plasmas e Fus\~ao Nuclear, Instituto Superior T\'ecnico, Universidade de Lisboa, 1049-001 Lisbon, Portugal
}%

\begin{document}

\maketitle

\begin{abstract}
We present a novel mechanism in which plasma electrons and ions optically acquire angular momentum during local pump depletion of an azimuthally polarized laser, despite the laser carrying none. Using theoretical considerations and multi-dimensional particle-in-cell simulations, we find that this process is enabled by a strong frequency downshift at the gradually eroding laser pulse front. We further show that the angular momentum gained by the plasma electrons is compensated by the ions and by the combined electromagnetic fields of the laser and nonlinear plasma wave. By varying key laser parameters such as phase, frequency, and polarization, we demonstrate that the transverse momentum of high-energy electrons can be effectively controlled.
\end{abstract}

\section{Introduction}

The transfer of optical angular momentum plays a critical role in diverse research areas, ranging from the precise control of nano-particle motion \citep{porfirev2023light} to the generation of substantial axial magnetic fields \citep{talin_inverse_1975,najmudin_measurements_2001,shi_magnetic_2018}, the development of compact accelerators for twisted electron beams with relativistic energies \citep{vieira_optical_2018}, and the emission of radiation with angular momentum from the spiral motion of high-energy electrons \citep{katoh2017angular, seipt2014structured, luis2019radiation}.

Optical angular momentum transfer in underdense plasmas has been studied in the context of the inverse Faraday effect, which was initially explored using circularly polarized lasers \citep{talin_inverse_1975, sheng_inverse_1996, gorbunov_magnetic_1998, frolov_excitation_2004, naseri_axial_2010, liu2023inverse, haines_generation_2001, kostyukov_inverse_2001, najmudin_measurements_2001}. Later studies included lasers with orbital angular momentum \citep{allen}, relying on the direct absorption of laser angular momentum \citep{ali_inverse_2010, nuter_gain_2020,longman_kilo-tesla_2021}.
Moreover, angular momentum transfer to plasma waves and energetic charged particles has been demonstrated using lasers with orbital angular momentum \citep{baumann2018electron}, spatio-temporally coupled beams forming light springs \citep{vieira_optical_2018}, and twisted beat waves \citep{saberi2017ponderomotive, shi_magnetic_2018}.

So far, optically induced plasma electron rotation has primarily been associated with circularly polarized lasers and lasers with orbital angular momentum. These lasers exhibit distinct types of angular momentum—spin and orbital—which enable the angular motion of electrons through direct absorption or higher-order nonlinearities \citep{haines_generation_2001, sheng_inverse_1996, frolov_excitation_2004}.
The time-averaged angular momentum density of an electromagnetic wave is given by \( \langle\mathcal{L}\rangle = 1/(4 \pi c)\vec{r} \times \langle \vec{E} \times \vec{B} \rangle \), where \( \vec{E} \) and \( \vec{B} \) are the electric and magnetic field vectors, expressed in Gaussian CGS units, and \( c \) is the speed of light.
A laser beam carries angular momentum when the time-averaged momentum density \( 1/(4 \pi c)\langle \vec{E} \times \vec{B} \rangle \) has an azimuthal component.

Here, we demonstrate that the plasma can gain angular momentum from lasers without angular momentum, such as azimuthally polarized lasers.
Azimuthally polarized laser pulses consist of two orthogonally polarized transverse electromagnetic (TEM\(_{01}\)) modes, i.e., \( \text{TEM}_{01 (x)} \hat{e}_y - \text{TEM}_{01 (y)} \hat{e}_x \) \citep{oron_formation_2000,nesterov_laser_2000}. At focus, the vector potential is described by \(\vec{A}(r) = a_0 (r/w_0) \exp\left(-r^2/w_0^2\right) \hat{e}_\theta\), where \( w_0 \) is the beam waist and $\hat e_\theta = -\sin \theta \hat e_x + \cos \theta \hat e_y$ is the unit vector in polar direction with $\theta = \tan^{-1} (y/x)$. This configuration exhibits a hollow intensity profile and corresponds to a transverse electric (TE) mode, featuring an electric field component, \( E_\theta \), oriented in polar direction and perpendicular to the propagation direction, with radial \( B_r \) and axial \( B_z \) components of the magnetic field.
As a result, its time-averaged momentum density is directed along the z-axis, \( 1/(4 \pi c)\langle \vec{E} \times \vec{B} \rangle = 1/(4 \pi c)\langle E_\theta B_r \rangle \hat{e}_z \), similar to radial polarization and linearly polarized Gaussian beams \citep{volke-sepulveda_orbital_2002}. Therefore, these lasers have zero angular momentum density and thus have not been previously associated with optically induced electron rotation.

Through analytical considerations and one- and three-dimensional particle-in-cell (PIC) simulations using OSIRIS \citep{osiris}, we show that plasma rotation arises from canonical momentum conservation and is facilitated by a non-vanishing vector potential that trails the main laser pulse, a phenomenon that develops during local pump depletion \citep{decker_evolution_1996}. We further demonstrate the conservation of total angular momentum and that the mechanism can be tuned, mainly through the laser polarization and the laser-to-plasma frequency ratio.

In the following, we normalize the electric \(\vec{E}\) and magnetic fields \(\vec{B}\) to \(m_e c \omega_p / e\). The vector potential \(A\) is normalized to \(m_e c / e\), where the dimensionless vector potential is given by \(a_0 = e A / m_e c\). Distances \(\vec{x}\), velocity \(v\), and time \(t\) are normalized to \(c / \omega_p\), \(c\), and \(1 / \omega_p\), respectively. The plasma frequency is defined as $\omega_p = \sqrt{4 \pi n_e e^2/m_e}$, where \(n_e\) is the background plasma density and \(m_e\) is the electron mass.

\section{One-dimensional analysis of local pump depletion and transverse momentum gain of electrons}

It is instructive to review the self-consistent laser-plasma dynamics in one dimension to study how the mechanism works. 
We simulate an ultra-intense short laser pulse with a dimensionless amplitude $a_0 = 6$ ($I \approx 8\times10^{19}$ W/cm$^2$) and FWHM pulse duration $\tau = 3 / \omega_p$ ($\sim$ 10 fs). The laser is linearly polarized (in $y$-direction) with a central frequency $\omega_0 = 8 \, \omega_p$. Initially, the laser pulse's vector potential $\vec A$ vanishes (nearly zero) outside the laser pulse. 
The underdense plasma starts at \(z = 34 \, c/\omega_p\) and is homogeneous with a density of \(n_e = 1/64 \, n_{cr} \approx 2.7 \times 10^{19} \, \text{cm}^{-3}\), where \(n_{cr} \approx 1.7 \times 10^{21} \, \text{cm}^{-3}\) corresponds to the critical density for a laser wavelength of \(\lambda = 800 \, \text{nm}\). The ions are assumed to be static.
The simulation is performed in a moving window that travels at the speed of light $c$. The simulation box is $68 \, c/\omega_p$ long, divided into $6800$ cells and one particle per cell. 

\begin{figure}
    \centering
    \includegraphics[width=1.0\linewidth]{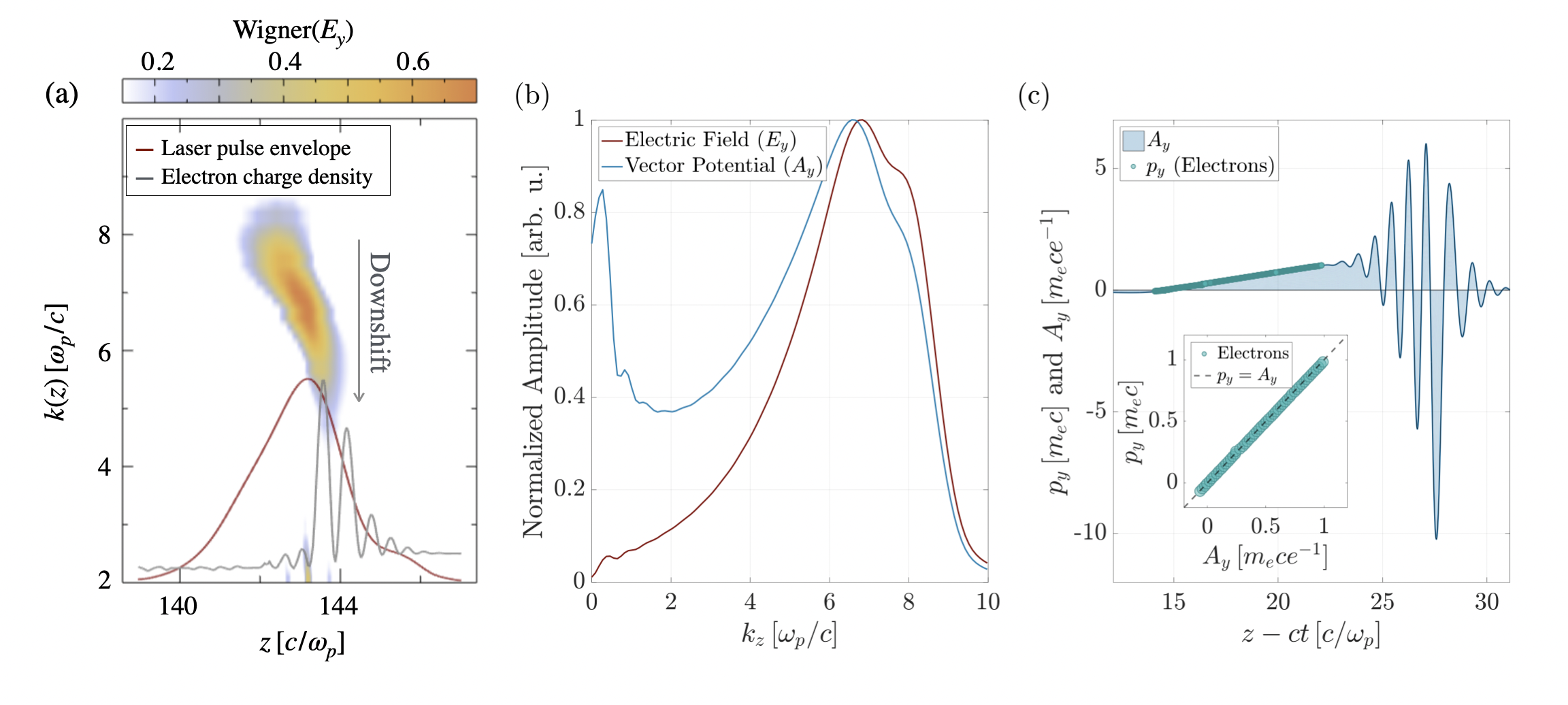}
    \caption{One-dimensional OSIRIS simulation results at time $t \approx 116/\omega_p$ showcase the frequency downshift of the laser electric field $E_y$, which manifests in the vector potential $A_y$ amid local pump depletion, enabling the plasma electrons to acquire transverse momentum in the nonlinear wakefield behind the main laser pulse. \textit{(a)} The frequency of the laser electric field (envelope of $E_y$ in red) drops at the leading edge (Wigner distribution of $E_y$), which resides in a steep electron charge density spike (grey line). \textit{(b)} The normalized amplitude of the low $k$-spectrum of the vector potential $A_y$ (blue line) surpasses that of the laser's electric field $E_y = -\partial A_y/\partial t$ (red line). \text{(c)} The vector potential $A_y$ (blue line) exhibits a long wavelength offset trailing the laser pulse ($(z-ct) \lesssim 25 \, c/\omega_p$ in the co-moving frame), where the electrons acquire transverse momentum,  following canonical momentum conservation $p_y = A_y$ (highlighted in the inset).}
    \label{fig:fig1}
\end{figure}

Figure \ref{fig:fig1} illustrates how local pump depletion enables the plasma electrons to acquire transverse momentum in a nonlinear wakefield behind the main laser pulse. This process is driven by a localized frequency downshift at the high-intensity laser pulse front (Fig. \ref{fig:fig1}(a)), the formation of a long-wavelength offset in the laser's vector potential (Fig. \ref{fig:fig1}(b)), which manipulates the electrons' transverse momentum through canonical momentum conservation (Fig. \ref{fig:fig1}(c)).

The laser pulse front pushes electrons away from regions of high intensity, thereby modifying the refractive index through electron density modulation and relativistic mass increase, $\eta = \sqrt{1-\omega_p^2/(\omega^2 \gamma)}$, where $\gamma = \sqrt{1+a^2/2}$ for a linearly polarized laser \citep{esarey1996overview}.
This interaction ensures that the laser front continuously encounters a positive gradient of the refractive index, $\partial \eta/ \partial \xi$, where $\xi = t-z/c$  is the speed of light variable. As a result, the laser frequency (and thus the axial wavenumber $k_z$) shifts down via phase modulation \citep{mendoncca1994regular,e1998kinetic, Silva.2001}, where the frequency change can be described by $ (1/\omega) (\partial \omega/\partial t) = - (1/c) (\eta^{-2}) (\partial \eta/\partial \xi)$ \citep{mori1997physics,tsung_generation_2002}. The plasma frequency $\omega_p$ marks the smallest frequency that can propagate.
This localized frequency downshift is observable in the Wigner transform of the laser's electric field $E_y$ in Fig. \ref{fig:fig1}(a). The position of the frequency downshift overlaps with the electron density spike (grey line) and the steepened leading edge of the laser pulse's electric field (envelope in red line) \citep{Silva.2001,vieira2010onset}.

The frequency downshift manifests in an increase of the vector potential, which acquires a substantial long-wavelength offset while the electric field appears eroded \citep{bulanov_nonlinear_1992, decker_evolution_1996, tsung_generation_2002}.
This is due to the conservation of the total wave action, $\mathcal{N} \sim \vert A_y \vert^2 \omega = \text{constant}$ \citep{bulanov_nonlinear_1992,tsung_generation_2002}; the vector potential increases while the laser frequency decreases.
As a result, the $k_z$-spectrum of the vector potential $A_y$ is increased for low wavenumbers, peaking at a few of the plasma wavenumber $k_p = \omega_p/c$ (see Fig. \ref{fig:fig1}(b)). 

The longer wavelength components travel with a smaller group velocity, where the group velocity can be approximated by $v_g \simeq c \left(1 - \omega_p^2/(2\omega^2)\right)$ \citep{mori1997physics}. Thus,  in Figure \ref{fig:fig1}(c), we observe a long wavelength offset trailing the main part of the laser pulse. 
The finite vector potential behind the laser pulse gradually decreases until it vanishes at the back of the wakefield. The electrons (blue dots) acquire a finite transverse momentum $p_y=A_y$, following the conservation of canonical momentum (see inset in Fig. \ref{fig:fig1}(c)).

This mechanism has been related to pulse compression \citep{e1998kinetic,tsung_generation_2002,gordon2003asymmetric,faure2005observation} , and proposed as a diagnostic for wakefields \citep{dias1998photon,murphy2006evidence,shiraishi2013laser,downer2018diagnostics}. 
In addition, the concept has been extended to explore the generation of mid-infrared to infrared pulses \citep{zhu2012studies,zhu2013pulsed,chen_high_2015,zhu2020efficient,pai2010generation,schreiber2010complete,nie2020photon} and betatron-like radiation from energetic electrons \citep{lu2021ultra}.

\begin{figure}
    \centering
    \includegraphics[width=0.9\linewidth]{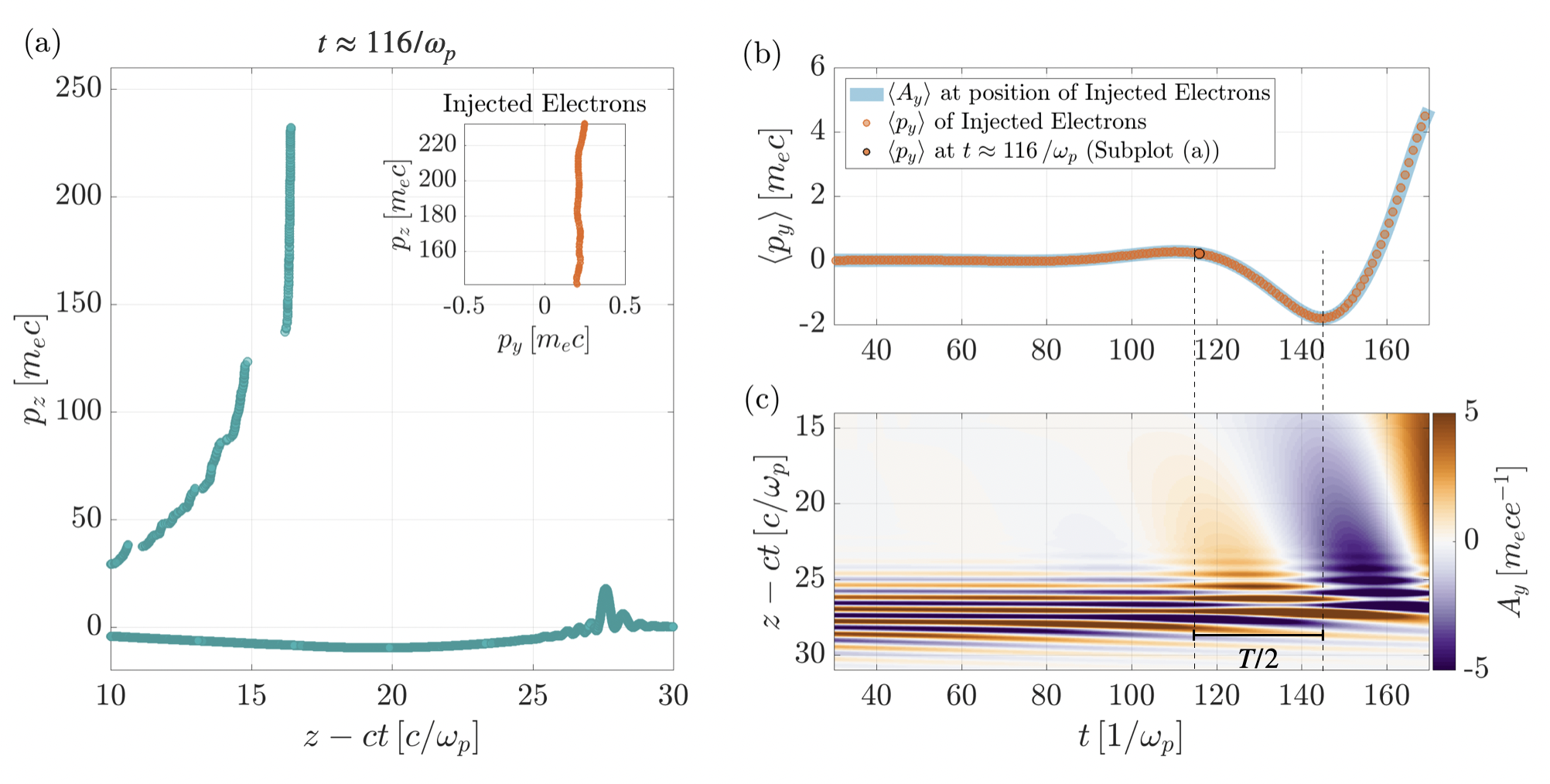}
    \caption{One-dimensional OSIRIS simulations show self-injected electron transverse momentum oscillations in a nonlinear wakefield, with increasing magnitude and a period linked to the laser's leading-edge erosion. \textit{(a)} The electrons' position, $z - ct$ in the co-moving frame, and longitudinal momentum, $p_z$, at time $\approx 116/\omega_p$, show high-energy electrons with $p_z > 140 \, m_e c$ concentrated at $(z-ct) \approx 16 \, c/\omega_p$, with their transverse momentum around $p_y \approx 0.25 \, m_e c$, as highlighted in the inset (orange dots). \textit{(b)} The time evolution of the mean transverse momentum of the high-energy electrons, $\langle p_y \rangle$, represented by orange dots, aligns well with the corresponding vector potential at their position, $\langle A_y \rangle$, shown as a blue line. The black-circled position corresponds to the time step of subplot \textit{(a)}. \textit{(c)} The time evolution of the vector potential $A_y$ in the co-moving frame illustrates the erosion of the leading edge and the development of an oscillating long-wavelength offset. The oscillation half-period, $T/2 \approx 25/\omega_p$, is highlighted by dashed lines, illustrating the connection between the erosion of the leading edge, the long-wavelength offset, and the transverse momentum of the electrons shown in \textit{(b)}.}
    \label{fig:fig2}
\end{figure}

In this setup, electrons can be efficiently accelerated due to the high accelerating gradients within the nonlinear wakefield, reaching gradients of GV/m \citep{esarey1996overview}. Within short propagation distances and times, self-injected electrons become trapped in the wakefield and are accelerated to high energies (\( \gtrsim 100 \, \text{MeV} \)).
In Fig. \ref{fig:fig2}(a), we demonstrate that in addition to their longitudinal acceleration, the high-energy electrons acquire transverse momentum. Specifically, the self-injected electrons with $p_z \gtrsim 140 \, m_e c$ localized around the same axial position $(z-ct) \approx 16 \, c/\omega_p$ within the nonlinear wakefield all contain similar transverse momentum $p_y \approx 0.25 \, m_e c$ (highlighted in the inset of Fig. \ref{fig:fig2}(a)).

Furthermore, during local pump depletion, the transverse momentum of the injected electrons oscillates with increasing magnitude \citep{nerush2009carrier}, matching the long wavelength offset of the vector potential, as shown in Fig. \ref{fig:fig2}(b) and (c).
The oscillation period of the long-wavelength offset of the vector potential is closely linked to the etching velocity—the rate at which the laser's leading edge erodes, defined as \( v_{\text{etch}} \approx c (\omega_p^2 / \omega_0^2 \)) \citep{decker_evolution_1996}. The oscillation period can be approximated as \( T \sim \lambda / v_{\text{etch}} \), where \( \lambda \) is the laser wavelength inside the plasma.
In Fig. \ref{fig:fig2}(c), we demonstrate an agreement between the oscillation of the long-wavelength offset and the erosion of the leading edge of the vector potential. The observed period of the oscillation, highlighted by dashed lines \( T/2 \approx 25 / \omega_p \), matches the theoretical estimate \( T/2 \approx (\lambda/2)\times \omega_0^2/(c \omega_p^2) \approx (\pi / 8) \times 64 / \omega_p \approx 25 / \omega_p \).

This section presents key results, demonstrating that the transverse momentum of plasma electrons is tunable according to the laser's polarization and through the laser-to-plasma frequency ratio, which governs the gradual erosion of the laser’s leading edge. This enhances the control of plasma electron dynamics through the wide range of accessible laser polarization states, e.g., linear, circular, radial, azimuthal, and even spatially varying polarization profiles.

\section{Three-dimensional analysis of angular momentum gain of electrons}

\begin{figure}
    \centering
    \includegraphics[width=1.0\linewidth]{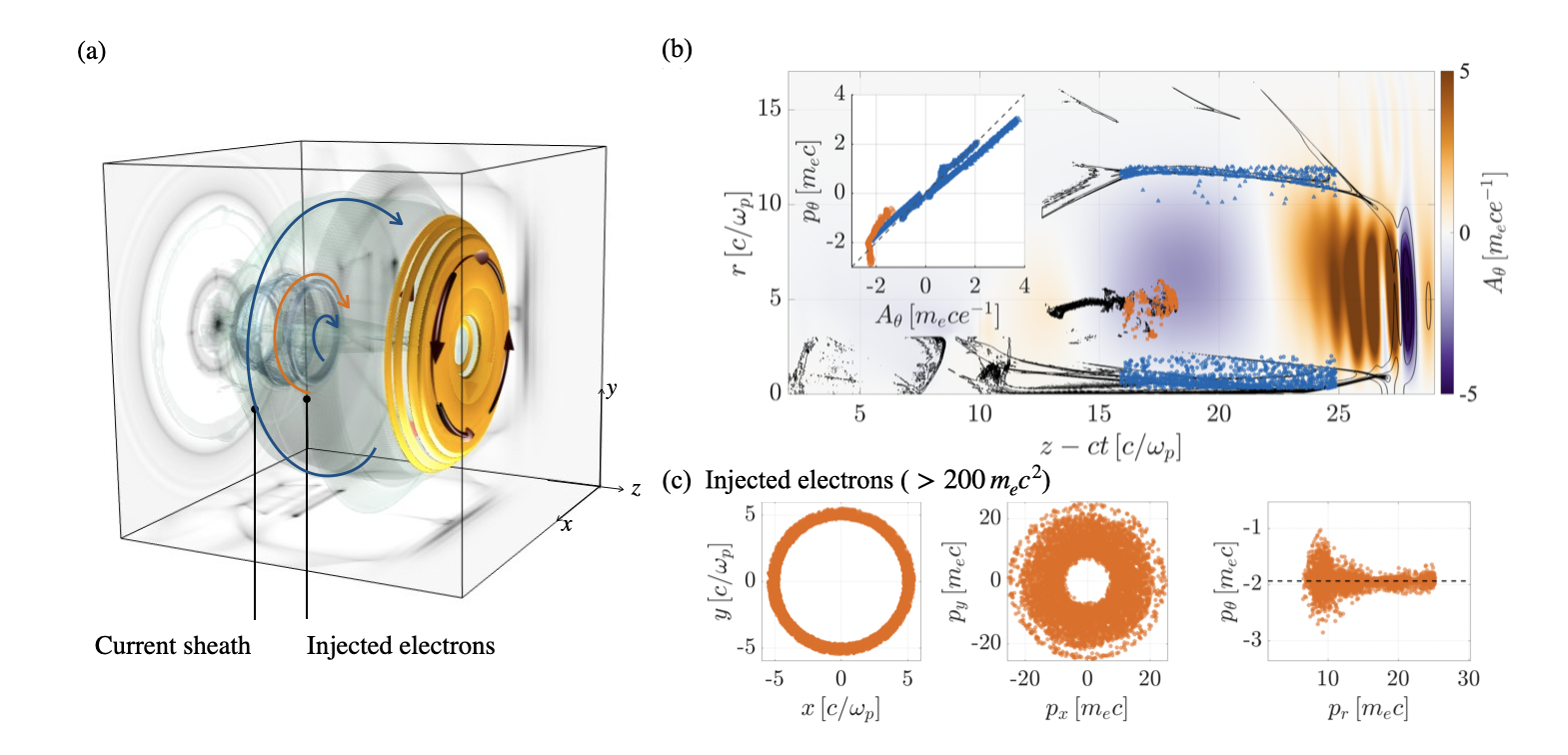}
    \caption{Three-dimensional OSIRIS simulation results at \( t \approx 150/\omega_p \) illustrate the angular momentum gain of plasma electrons in a nonlinear wakefield, driven by an azimuthally polarized laser pulse -- facilitated by the development of a long-wavelength offset in the laser's azimuthal vector potential due to local pump depletion.
\textit{(a)} The hollow intensity laser pulse (orange isosurfaces) with azimuthal polarization (black arrows) drives a donut-shaped nonlinear wakefield in the bubble regime (gray isosurfaces). Projections show the spatial distribution of the electron charge density, highlighting increased density at the laser pulse front. The blue and orange arrows indicate the direction of rotation of the electrons belonging to the electron sheath composing the donut-shaped bubble (blue) and the injected electrons inside the bubble (orange). The illustrated box dimensions are \( 20 \times 20 \times 20 \, (c/\omega_p)^3 \).
\textit{(b)} The vector potential \( A_\theta \) (yellow-purple colormap) exhibits a long-wavelength offset trailing the laser pulse (\( z \lesssim 25 \, c/\omega_p \)). Electrons in the inner and outer current sheaths (blue dots) and the self-injected electrons (orange dots) acquire azimuthal momentum following canonical momentum conservation, \( p_\theta = A_\theta \), (see inset).
\textit{(c)} The high-energy (\( >200 \, m_e c^2 \)) injected electrons form a ring in both configuration and transverse momentum space, exhibiting a well-defined angular momentum with a mean value of $\langle p_\theta \rangle \approx -2 m_e c$ (dashed line).}
    \label{fig:fig3}
\end{figure}

Building on our one-dimensional analysis, we extend our study to three dimensions, confirming the robustness of electron transverse momentum gain. In this context, we show that plasma electrons acquire angular momentum from a high-intensity azimuthally polarized laser pulse in underdense plasma, driving a nonlinear wakefield in the bubble regime \citep{pukhov2002laser}. Figure~\ref{fig:fig3} shows the angular momentum of electrons from the bubble's current sheath and injected electrons.  

We simulate an azimuthally polarized laser pulse with a dimensionless amplitude of $a_0 = 6$, FWHM pulse duration $\tau = 3 / \omega_p$, a central frequency $\omega_0 = 8 \, \omega_p$, and a laser spot size  $w_0 = 5.5 \,c/\omega_p$. Considering a wavelength of $\lambda = 800$ nm, this corresponds to a laser with about $600\, \text{mJ}$,  $w_0 = 5.6 \, \mu$m, and with pulse duration $\tau \approx 10$ fs propagating in a plasma with density $n_e = 2.7\times10^{19}/\text{cm}^3$. The simulation is performed in a moving window that travels at the speed of light $c$. The simulation box has the dimensions $34\times34\times34 \, (c/\omega_p)^3$, divided into $1700\times1700\times1700$ cells and 4 particles per cell.

In Fig. \ref{fig:fig3}(a), we present simulation results of an azimuthally polarized laser pulse driving a donut-shaped plasma bubble, reflecting the laser’s hollow intensity distribution \citep{vieira_nonlinear_2014}. The electrons composing the bubble and the self-injected electrons have angular momentum, as indicated by the arrows in blue and orange.
Figure \ref{fig:fig3}(b) shows the driver’s vector potential at \( t \approx 150 / \omega_p \), undergoing local pump depletion and developing a substantial long-wavelength offset that remains confined within the bubble structure. The electrons from the inner and outer current sheaths (highlighted in blue) and the high-energy self-injected electrons (highlighted in orange) possess angular momentum in agreement with canonical momentum conservation, \( p_\theta = e A_\theta \) (see inset in Fig. \ref{fig:fig3}(b)). Here, $p_\theta =  m_e \gamma r \dot \theta$ is the angular momentum, where $\gamma = \sqrt{1 + \vert \vec p \vert^2}$ is the relativistic factor, $\dot \theta$ is the electron's angular frequency, and $A_\theta$ is the laser's azimuthal vector potential at the position of the electrons (see more on angular momentum conservation in Appendix \ref{section:angular}). 
In Fig. \ref{fig:fig3}(c), we highlight the high-energy (\( >200 \, m_e c^2 \)) self-injected electrons, forming a ring structure in physical and transverse momentum space. The electrons carry a well-defined angular momentum, with a mean value of \( \langle p_\theta \rangle \approx -2 m_e c \). The total charge of these high-energy electrons is  \( \approx 212 \, \mathrm{pC} \).

Consistent with angular momentum conservation, the angular momentum gained by the plasma electrons is compensated by the ions and by the combined electromagnetic fields of the laser and the nonlinear plasma wave. To show this, we consider the integral form of the angular momentum conservation law for a closed system, without particle loss, given by \citep{jackson}
\begin{equation}
    \int_V \left( \mathcal{L}_\mathrm{mech} + \mathcal{L}_\mathrm{field} \right) \, d^3x + \int \left( \int_S \hat{n} \cdot \mathbf{M} \, da \right) dt = 0,
    \label{Eq:totalAngular}
\end{equation}
where \( \mathcal{L}_\mathrm{mech} = n_e \vec{r} \times \vec{p_e} + n_i \vec{r} \times \vec{p_i}\) is the mechanical angular momentum density of the electrons and ions, and \( \mathcal{L}_\mathrm{field} = 1/(4 \pi c) \vec r \times (\vec E \times \vec B) \) represents the angular momentum density carried by the fields. Here, the electric and magnetic fields are the combined wakefields and laser fields. The term \( \mathbf{M} = \vec{r} \times \mathbf{T} \) stands for the angular momentum flux, with \( \mathbf{T} \) being the Maxwell stress tensor normal to the closed surface $S$. 

\begin{figure}
    \centering
    \includegraphics[width=0.9\linewidth]{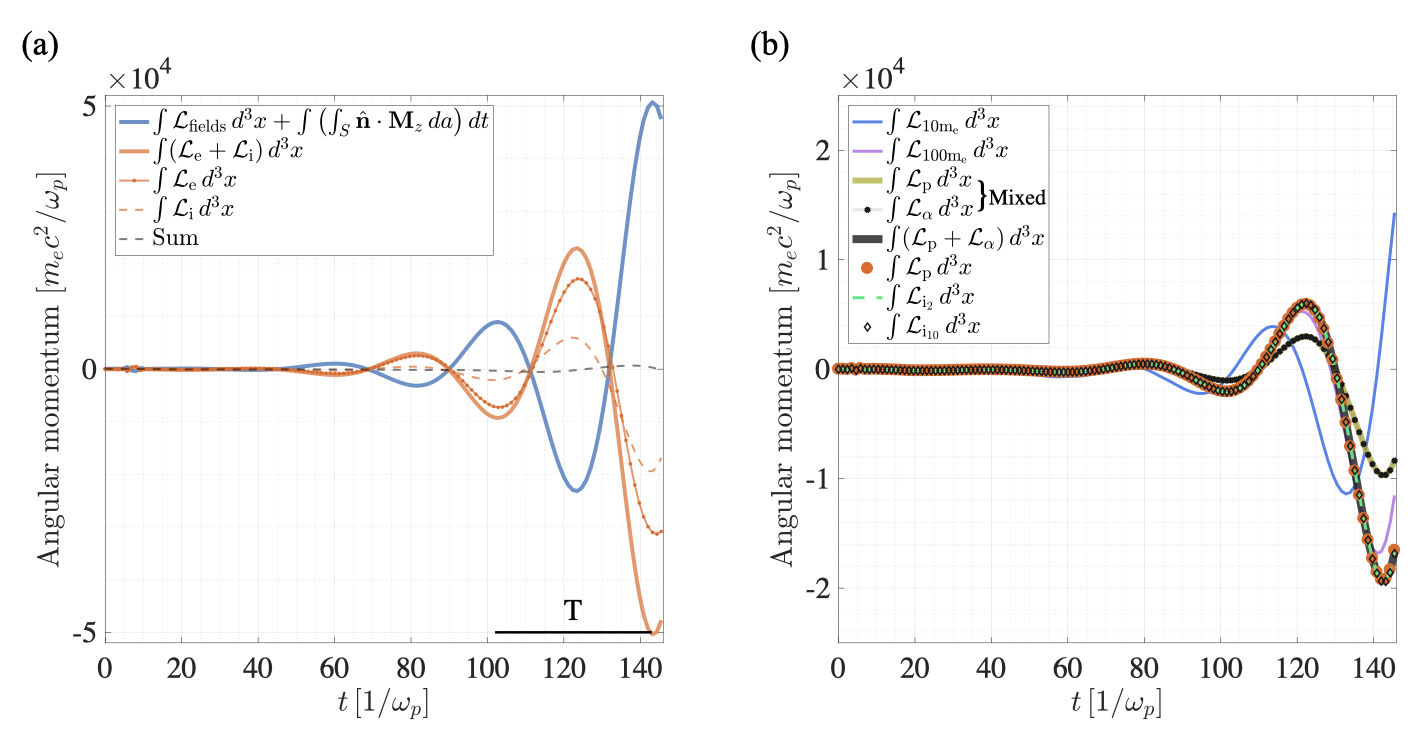}
  \caption{Three-dimensional OSIRIS simulation results showing the evolution of angular momentum components, as defined in Eq.~\ref{Eq:totalAngular}. 
\textit{(a)} Angular momentum gained by the plasma electrons and ions (orange line) is compensated by that of the combined wakefield and laser fields (blue line). Individual contributions from electrons and ions are indicated in orange by crosses and a dashed line. The horizontal black line labeled ``T'' indicates the oscillation period. 
\textit{(b)} The angular momentum acquired by protons and heavier ions is independent of their mass. This trend deviates for lighter species, as their motion influences local pump depletion and wakefield formation. We compare positively charged species with ten (blue line) and one hundred (purple line) times the electron mass, a mixed gas of protons (green line) and alpha particles (grey-dotted) and their sum (black line), protons (orange-dotted), ions with twice the proton mass (green dashed line), and ions with ten times the proton mass (black diamonds).}
    \label{fig:newfig}
\end{figure}

Figure~\ref{fig:newfig}(a) illustrates the individual and combined angular momentum contributions from the electrons, the ions, and the electromagnetic fields, as expressed in Eq.~\ref{Eq:totalAngular} (see Movie 1 in the supplementary material for the full evolution). The mechanical angular momentum of the electrons and ions is balanced by the angular momentum of the electromagnetic fields. The electrons’ angular momentum exceeds that of the ions, which can be attributed to electron accumulation in regions of enhanced vector potential, where they naturally acquire more angular momentum. Figure~\ref{fig:newfig}(b) shows that the angular momentum acquired by ions is mass independent, consistent with canonical momentum conservation, $p_\theta = -e A_\theta$. This is evident from the overlapping curves for the protons, the heavier ions, and the proton–alpha mixture, with the mixture’s angular momentum dividing evenly between the two species. Only the curves for the lighter positively charged species ($10\,m_e$ and $100\,m_e$) deviate from this trend, as their motion influences local pump depletion and wakefield formation.
The period of the angular momentum oscillation can be approximated as \( T \sim \lambda / v_{\text{etch}} \), similar to the one-dimensional discussion around Fig.~\ref{fig:fig2}(b) and (c). To our knowledge, this is the first quantitative verification of angular momentum conservation in a PIC simulation.

The mechanism is robust and tunable, primarily governed by the laser-to-plasma frequency ratio and the laser intensity, which must be sufficiently high (\( a_0 > 3 \), \citep{tsung_generation_2002}) for local pump depletion to occur. Other influencing factors include the laser phase, polarization, chirp, and pulse duration.

\begin{figure}
    \centering
    \includegraphics[width=0.7\linewidth]{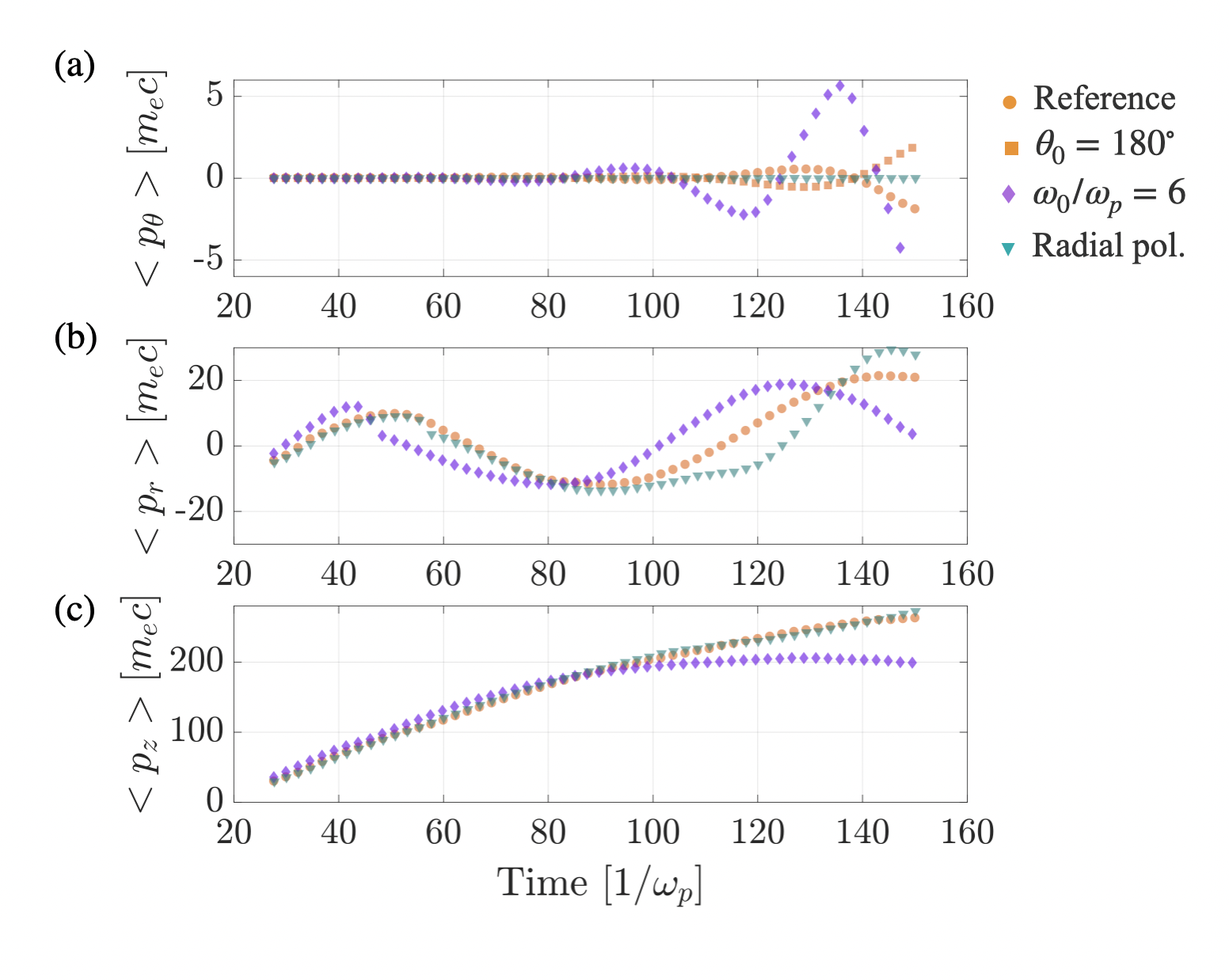}
    \caption{Three-dimensional OSIRIS simulation results demonstrate how adjustments of laser parameters allow for the control of angular (transverse) momentum of high-energy electrons (\( > 80\% \) of the maximum energy) under local pump depletion. \textit{(a)} to \textit{(c)} show the time evolution of the mean azimuthal momentum \( \langle p_\theta \rangle \), mean radial momentum \( \langle p_r \rangle \), and mean longitudinal momentum \( \langle p_z \rangle \), respectively, for different laser and plasma parameters. 
    Orange dots correspond to the reference case with $\theta_0 = 0$, $\omega_0/\omega_p = 8$ and azimuthal polarization; 
orange squares correspond to a change in the initial laser phase, \( \theta_0 = 180^\circ \); purple diamonds represent a change in the laser-to-plasma frequency ratio, \( \omega_0 / \omega_p = 6 \); and blue triangles illustrate the transition to radial polarization.
Note that the phase change affects only $p_\theta$, such that the points overlap in \textit{(b)} and \textit{(c)}.
}
    \label{fig:fig4}
\end{figure}

Figure \ref{fig:fig4} shows a parameter scan on the tunability of angular (transverse) momentum under local pump depletion, focusing on the momentum space evolution of self-injected high-energy electrons. 
We examine the effect of three key parameters on the high-energy electrons (\( > 80\% \) of the maximum energy) - the initial phase of the laser, which we vary from \( \theta_0 = 0 \) to \( 180^\circ \) (depicted by orange squares in Fig. \ref{fig:fig4}), the laser-to-plasma frequency ratio, which we vary from \( \omega_0 / \omega_p = 8 \) to \( 6 \) (depicted by purple diamonds), and the laser polarization where we change from azimuthal to radial polarization (depicted by blue triangles). Radial polarization is composed as \( \text{TEM}_{01 (x)} \hat{e}_x + \text{TEM}_{01 (y)} \hat{e}_y \) \citep{oron_formation_2000,nesterov_laser_2000}.

Figure~\ref{fig:fig4}(a) shows the evolution of the mean electron angular momentum \( \langle p_\theta \rangle \), which exhibits oscillations with increasing amplitude, consistent with the etching-driven dynamics discussed around Figs.~\ref{fig:fig2} and~\ref{fig:newfig}.
Given the multidimensional structure and transverse focusing fields, the electrons also undergo betatron oscillations with a frequency \( \omega_\beta = \omega_p / \sqrt{2 \gamma} \)  \citep{tajima1979laser,esarey2009physics}, as observed in the radial momentum \( \langle p_r \rangle \) in Fig. \ref{fig:fig4}(b). The longitudinal momentum \( \langle p_z \rangle \) increases steadily before eventually saturating, as shown in Fig. \ref{fig:fig4}(c).
Adjustments to laser parameters influence the polarity, magnitude, and distribution of angular (transverse) momentum of the injected electrons.
In Fig. \ref{fig:fig4}(a), we see that varying the laser's initial phase influences the polarity of the angular momentum \( \langle p_\theta \rangle \), but has no influence on the radial and axial momentum. Reducing the laser-to-plasma frequency ratio impacts the oscillation period and magnitude of the angular momentum, due to the increased etching velocity \( v_{\text{etch}} \approx c(\omega_p^2 / \omega_0^2) \). It also allows for increased angular momentum, but the longitudinal momentum saturates more quickly. This can be seen in Fig. \ref{fig:fig4}(c). 
On the other hand, radial polarization does not lead to the angular motion but introduces significant modifications to the radial momentum \( \langle p_r \rangle \), adding a distinct modulation on top of the betatron oscillations, which becomes prominent as pump depletion progresses (\( t > 120/\omega_p \)), as shown in Fig. \ref{fig:fig4}(b).
Experimental diagnostics of this mechanism could rely on the rotating high-energy electrons, which emit betatron radiation with spatial or polarization features that reflect their angular momentum~\citep{luis2019radiation}. While this effect was not modeled in our simulations, the resulting X-rays could be measured using standard betatron radiation diagnostics~\citep{phuoc2006imaging,thaury2013angular,corde2013observation}.

Additionally, the electromagnetic field structure from the downshifted azimuthally polarized laser fields, particularly \( E_\theta \), \( B_r \), and \( B_z \), inside the bubble is orthogonal to that of conventional wakefields, offering a measurable signature, for example via electron probing~\citep{zhang2016capturing,raj2020probing,wan2024real}.

\section{Conclusions}

In conclusion, we demonstrated that plasma electrons and ions can acquire substantial angular momentum via canonical momentum conservation, driven by a high-intensity azimuthally polarized laser pulse undergoing local pump depletion in underdense plasma. Through a combination of analytical considerations and particle-in-cell simulations, we explored the mechanism of transverse momentum gain of electrons in one-dimensional scenarios and extended our analysis to three dimensions, obtaining consistent results.
This mechanism relies on local pump depletion of the laser pulse, which facilitates the development of a long-wavelength offset in the vector potential trailing the pulse. This offset enables the rotation of the wakefield's sheath currents and the self-injected electrons. The highest-energy electrons exhibit angular momentum with a small spread and undergo characteristic oscillations linked to the erosion of the laser pulse front.
We found that the angular momentum gained by the electrons is compensated by that of the ions and the combined fields of the laser and the nonlinear plasma wave.
Finally, we highlighted the controllability of angular and transverse momentum gain under local pump depletion by varying key laser parameters, including the initial phase, laser-to-plasma frequency ratio, and polarization. 
A more systematic parametric study, including variations in laser pulse shape and intensity and electron density, would help further characterize the robustness and scaling of this mechanism regarding potential experimental realization, and is left for future work.

Future work will also aim to test the limits of this scheme, explore exotic laser polarizations, investigate radiation emitted by high-energy electrons, and incorporate ionization effects.
Finally, the presented mechanism potentially impacts setups involving lasers with angular momentum, such as circularly polarized Gaussian and Laguerre-Gaussian beams, which merits further investigation.



\section*{Acknowledgments}

We would like to acknowledge the anonymous reviewer for their valuable and constructive comments, which helped us to improve the quality of this work.
The authors used AI tools to assist with copy editing (improvements on readability and style, and to ensure that the texts are free of errors in grammar, spelling, and punctuation).

\section*{Funding}
We gratefully acknowledge EuroHPC for awarding us access to LUMI-C at CSC (Finland) and Rede Nacional de Computação Avançada (RNCA) for access to Marenostrum V at the Barcelona Supercomputer Center.
This work was supported by the EuPRAXIA-PP under grant agreement No. 101079773 and the FCT (Portugal) Foundation for Science and Technology under Project No. X-MASER-2022.02230.PTDC, FCT PD/ BD/142971/2018, UID/FIS/50010/2020, UI/BD/151559/2021, and IPFN-CEEC-INST-LA3/IST-ID. This work was also supported by a fellowship from the “la Caixa” Foundation (ID 100010434), under fellowship code LCF/BQ/DI19/11730025.

\section*{Declaration of interests}

The authors report no conflict of interest.

\appendix

\section{Angular momentum conservation}\label{section:angular}

The Lagrangian for an electron in an electromagnetic wave in a wakefield can be written as \citep{jackson}
\begin{equation}
    L = -m_e c^2 \left(1 - \frac{\vec{v} \cdot \vec{v}}{c^2} \right)^{1/2} + \frac{e}{c} \vec{v} \cdot \vec{A} - e \phi,
\end{equation}
where \(\vec{v}\) is the velocity vector, \(c\) is the speed of light, \(\vec{A}\) is the laser's vector potential, \(\phi\) is the scalar potential (here the wakefield potential), and \(m_e, e\) are the electron's mass and charge, respectively.
Our system is cylindrically symmetric, and for this purpose, we write the Lagrangian in cylindrical coordinates as
\begin{equation}
    L = -m_e c^2 \left( 1-\frac{\dot z^2}{c^2}-\frac{r^2 \dot \theta^2}{c^2}-\frac{\dot r^2}{c^2}\right)^{1/2} - e \left( \dot z A_z + r \dot \theta A_\theta + \dot r A_r \right) + e \phi,
\end{equation}
where \(\dot z = dz/dt\) is the axial velocity, \(\dot \theta = d\theta/dt\) is the angular velocity, and \(\dot r = dr/dt\) is the radial velocity.
To determine whether angular momentum is conserved in this system, we calculate the generalized angular momentum
\begin{equation}
    P_\theta = \frac{\partial L}{\partial \dot \theta} = m_e \gamma r^2 \dot \theta -e A_\theta r,
\end{equation}
where \(\gamma = 1/\sqrt{1-(\dot z^2 + r^2 \dot \theta^2 + \dot r^2)/c^2}\). Here, $ m_e \gamma r \dot \theta = p_\theta$ is the azimuthal momentum, and there is no angular velocity dependence on the wakefield potential, i.e., \(\partial \phi /\partial \dot \theta = 0\).
The generalized angular momentum is conserved when the Hamiltonian, given by
\begin{equation}
    H = \sqrt{\left(c \vec P - e \vec A  \right)^2 + m_e^2 c^4} + e\phi,
\end{equation}
has no angular dependence, i.e.,
\begin{equation}
    \frac{d P_\theta}{dt} = -\frac{\partial H}{\partial \theta} = 0.
\end{equation}
This results in $P_\theta = 0$, because there is no initial angular momentum in the system, such that $p_\theta = e A_\theta$.

This condition holds for a cylindrical vector beam with azimuthal polarization, where the laser pulse has no angular dependence and the vector potential has only an azimuthal component, \(\vec A = A_\theta(r,z) \hat{e}_\theta\), where at focus $A(r,0)_\theta = a_0 (r/w_0) \exp\left(-r^2/w_0^2\right)$.
In this case, plasma electrons can only acquire angular momentum through canonical momentum conservation, $p_\theta = e A_\theta$.

\bibliographystyle{jpp}


\end{document}